\documentclass[twocolumn,aps,preprintnumbers,amsfonts,amsmath,amssymb,superscriptaddress,floatfix,10pt]{revtex4-2}

\usepackage{siunitx}
\usepackage{bm}
\usepackage{accents}
\usepackage{graphicx}
\usepackage[hidelinks]{hyperref}

\newcommand\figref[1]{Fig.~\!\ref{fig:#1}}
\newcommand\figureref[1]{Figure~\ref{fig:#1}}

\begin{document}
\title{
Experimental implementation of continuous-variable QAOA on a quad-rail lattice cluster state
}

\author{Shota~Yokoyama}
\email{Contact author: s.yokoyama@riken.jp}
\affiliation{RIKEN Center for Quantum Computing, 2-1 Hirosawa, Wako, Saitama 351-0198, Japan}

\author{Atsushi~Sakaguchi}
\affiliation{RIKEN Center for Quantum Computing, 2-1 Hirosawa, Wako, Saitama 351-0198, Japan}

\author{Jun-ichi~Yoshikawa}
\affiliation{RIKEN Center for Quantum Computing, 2-1 Hirosawa, Wako, Saitama 351-0198, Japan}

\author{Hironari~Nagayoshi}
\affiliation{Department of Applied Physics, School of Engineering, The University of Tokyo, 7-3-1 Hongo, Bunkyo-ku, Tokyo 113-8656, Japan}

\author{Warit~Asavanant}
\affiliation{RIKEN Center for Quantum Computing, 2-1 Hirosawa, Wako, Saitama 351-0198, Japan}
\affiliation{Department of Applied Physics, School of Engineering, The University of Tokyo, 7-3-1 Hongo, Bunkyo-ku, Tokyo 113-8656, Japan}
\affiliation{OptQC Corp., 3-28-13 Nishi-Ikebukuro, Toshima-ku, Tokyo 171-0021, Japan}

\author{Kan~Takase}
\affiliation{RIKEN Center for Quantum Computing, 2-1 Hirosawa, Wako, Saitama 351-0198, Japan}
\affiliation{Department of Applied Physics, School of Engineering, The University of Tokyo, 7-3-1 Hongo, Bunkyo-ku, Tokyo 113-8656, Japan}
\affiliation{OptQC Corp., 3-28-13 Nishi-Ikebukuro, Toshima-ku, Tokyo 171-0021, Japan}

\author{Takuji~Hiraoka}
\affiliation{OptQC Corp., 3-28-13 Nishi-Ikebukuro, Toshima-ku, Tokyo 171-0021, Japan}
\affiliation{Fixstars Amplify Corporation, 3-1-1 Shibaura, Minato-ku, Tokyo 108-0023, Japan}

\author{Akira~Furusawa}
\email{Contact author: akiraf@ap.t.u-tokyo.ac.jp}
\affiliation{RIKEN Center for Quantum Computing, 2-1 Hirosawa, Wako, Saitama 351-0198, Japan}
\affiliation{Department of Applied Physics, School of Engineering, The University of Tokyo, 7-3-1 Hongo, Bunkyo-ku, Tokyo 113-8656, Japan}

\author{Hidehiro~Yonezawa}
\email{Contact author: h.yonezawa@riken.jp}
\affiliation{RIKEN Center for Quantum Computing, 2-1 Hirosawa, Wako, Saitama 351-0198, Japan}

%%%%%%%%%%%%%%%%%%%%%%%%%%%%%%%%%%%%%%%%%%%%%%%%

\begin{abstract}
\noindent
We experimentally demonstrate the continuous-variable quantum approximate optimization algorithm (CV-QAOA) for multi-variable problems and multiple QAOA depths using a measurement-based CV quantum computing platform on a quad-rail lattice (QRL) cluster state.
We propose a systematic method to map arbitrary quadratic cost functions onto the QRL architecture and examine the resulting construction in settings involving up to 100 modes.
Using the programmable platform, we prepare the CV-QAOA ansatz and optimize the variational parameters via Bayesian optimization.
We then investigate the performance on quadratic optimization problems and observe that increasing the depth from 1 to 2 improves performance, whereas further increases yield only limited gains.
In contrast, numerical simulations under idealized conditions, assuming an infinite number of measurement shots and gradient-based optimization, indicate that the performance of CV-QAOA can improve with increasing depth, suggesting that the experimentally observed limitations primarily arise from noise accumulation and classical optimization challenges.
This work provides an experimental demonstration of CV-QAOA on a programmable CV platform and establishes a foundation for future developments of variational quantum algorithms in CV systems.
\end{abstract}

%%%%%%%%%%%%%%%%%%%%%%%%%%%%%%%%%%%%%%%%%%%%%%%%

\maketitle

% Introduction
\section{Introduction}
Quantum computers are expected to provide a computational platform that can efficiently solve problems that are difficult for classical computers~\cite{nielsen2010quantum}. Their potential applications have been discussed in a wide range of fields, including artificial intelligence~\cite{biamonte2017quantum}, quantum chemistry and drug discovery~\cite{cao2019quantum}, and finance~\cite{orus2019quantum}. Toward the realization of fault-tolerant quantum computation (FTQC), significant efforts have been made to develop quantum hardware based on various physical platforms, such as superconducting circuits~\cite{arute2019quantum}, trapped ions~\cite{monroe2021programmable}, neutral atoms~\cite{ebadi2021quantum}, and photonic systems~\cite{zhong2020quantum}. Both experimental and theoretical progress over the past decade has been remarkable.

At present, noisy intermediate-scale quantum (NISQ) devices~\cite{preskill2018quantum} are widely available, and the development of quantum algorithms that can run on such devices has been rapidly progressing. Understanding the practical capabilities of quantum computation in the NISQ regime requires investigations from both algorithmic and hardware perspectives.

Variational quantum algorithms (VQAs)~\cite{cerezo2021variational}, which combine shallow quantum circuits with classical optimization, have attracted much attention as a framework that can operate under the constraints of NISQ devices~\cite{kandala2017hardware}. In VQAs, a parameterized quantum circuit (ansatz) is treated as a black-box function, and its parameters are optimized using well-established classical optimization methods to achieve a desired task. This hybrid quantum-classical approach allows us to tackle practical problems with NISQ devices, driving the rapid development of VQAs.

A representative example of VQAs is the quantum approximate optimization algorithm (QAOA)~\cite{farhi2014quantum}, which has been applied to a wide range of problems, including combinatorial optimization~\cite{harrigan2021quantum}, portfolio optimization~\cite{hodson2019portfolio}, and related applications~\cite{blekos2024review}. In QAOA, increasing the depth, i.e., the number of repeated QAOA layers, generally enhances the expressibility of the ansatz, achieving a better approximation of an adiabatic evolution. However, deeper circuits are more sensitive to noise, and increasing the number of variational parameters makes classical optimization more difficult~\cite{zhou2020quantum, larocca2025barren}. To address these issues, prior work has explored error mitigation and several QAOA variants, including ma-QAOA~\cite{herrman2022multi} and FALQON~\cite{magann2022feedback}.

QAOA has also been theoretically proposed in continuous-variable (CV) systems~\cite{verdon2019quantum}, where continuous degrees of freedom are utilized instead of discrete qubits. Among various physical implementations of CV quantum computing, photonic systems are particularly promising due to their ability to generate large-scale entanglement deterministically, high-efficiency detection, and scalability via time-domain multiplexing~\cite{asavanant2022optical}. In recent years, measurement-based CV quantum processors have been experimentally realized~\cite{asavanant2019generation, larsen2019deterministic, larsen2021deterministic}, and cloud-based platforms with programmable multi-input and multi-step operations have started to emerge~\cite{yokoyama2025full}. These developments suggest that CV systems are also entering the NISQ regime.

Algorithmic implementations in CV quantum computing, however, have so far been limited to small-scale examples, and systematic demonstrations of NISQ-oriented CV quantum algorithms remain at an early stage. For instance, while proof-of-principle demonstrations for single-variable optimization problems have been reported~\cite{enomoto2023continuous}, the performance of CV-QAOA with increased ansatz expressibility or larger problem sizes has not yet been subjected to systematic experimental investigation. Therefore, evaluating scalability, ansatz expressibility, and robustness against noise in practical systems is highly important for further developing these algorithms.

In this work, we implement CV-QAOA on a programmable CV quantum processor and experimentally demonstrate its operation on multi-variable problems at multiple QAOA depths. Specifically, we consider a quadratic cost function over continuous variables and study the performance of CV-QAOA as a function of the QAOA depth and the number of variables. Both numerical simulations and experimental results are presented to evaluate the scaling behavior and performance limitations of CV-QAOA under realistic conditions. Although this specific implementation does not offer a quantum advantage over classical methods, this work provides a concrete implementation of a NISQ quantum algorithm in a CV system. As such, it establishes a foundational framework for future research and represents a step toward extending the VQA framework established in DV systems to CV quantum computing.

The remainder of this paper is organized as follows. In Sec.~\ref{sec:background}, we review the theoretical framework of CV-QAOA and the implementation of Gaussian operations on a quad-rail lattice (QRL) cluster state. In Sec.~\ref{sec:setup}, we describe the implementation of the CV-QAOA ansatz using measurement-based quantum computing (MBQC) on a QRL cluster state. In Sec.~\ref{sec:results}, we present experimental and numerical results, including performance analysis for quadratic optimization problems and their scaling with system size. Finally, Sec.~\ref{sec:conclusion} concludes the paper.

\section{Background}
\label{sec:background}
\begin{figure}[t!]\centering
\includegraphics[width=\columnwidth]{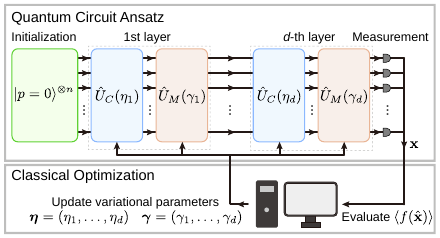}
\caption{
Schematic of the CV-QAOA algorithm.
The depth $d$ denotes the total number of QAOA layers, each comprising a cost operation $\hat{U}_C$ and a mixer operation $\hat{U}_M$.
}
\label{fig:1}
\end{figure}

In this paper, we consider CV quantum systems described by the $j$-th bosonic modes $\hat{a}_j$ and position and momentum quadrature operators
$\hat{x}_j = \frac{1}{\sqrt{2}} (\hat{a}_j + \hat{a}_j^\dagger)$ and
$\hat{p}_j = \frac{1}{i\sqrt{2}} (\hat{a}_j - \hat{a}_j^\dagger)$,
which satisfy $[\hat{x}_j, \hat{p}_{j'}] = i \delta_{j,j'}$ with $\hbar = 1$.

\subsection{CV-QAOA}
\label{sec:CV-QAOA}
Figure~\ref{fig:1} shows an overview of CV-QAOA~\cite{verdon2019quantum}. The goal of CV-QAOA is to solve optimization problems by finding a quantum state that minimizes a given cost function. This cost function is a continuous multivariable function over real numbers, typically defined as the expectation value of a cost Hamiltonian $\langle \hat{H}_C \rangle$. To achieve this, CV-QAOA generates a parametrized quantum state by alternately applying unitary operations generated by a cost Hamiltonian $\hat{H}_C$ and a mixer Hamiltonian $\hat{H}_M$, satisfying $[\hat{H}_C, \hat{H}_M] \neq 0$. An initial state is a momentum eigenstate $|p = 0 \rangle^{\otimes n}$, where $n$ is the number of modes. The generated quantum state is given by
\begin{align}
|\psi(\boldsymbol{\eta}, \boldsymbol{\gamma})\rangle
= \left(\prod_{k=1}^{d} \hat U_M(\gamma_k) \hat U_C(\eta_k) \right) |p = 0 \rangle^{\otimes n},
\end{align}
where $\hat U_C(\eta) = e^{-i \eta \hat H_C}$, $\hat U_M(\gamma) = e^{-i \gamma \hat H_M}$, $\boldsymbol{\eta} = (\eta_1, \dots, \eta_d)$ and $\boldsymbol{\gamma} = (\gamma_1, \dots, \gamma_d)$ are variational parameters, and $d$ is the QAOA depth, corresponding to the number of repeated QAOA layers. The parameters are optimized via a classical optimization process, where they are iteratively updated based on measurement outcomes of the generated state.

The mixer Hamiltonian can be chosen arbitrarily as long as it does not commute with the cost Hamiltonian. Typically, the cost Hamiltonian is defined in terms of position quadratures, $\hat{H}_C = f(\hat{\mathbf{x}})$, and the mixer Hamiltonian is chosen as a quadratic function of momentum, $\hat{H}_M = \sum_j \frac{1}{2} \hat{p}_j^2$. In this setting, the update rule for the $k$-th layer can be written as
\begin{align}
\hat{\mathbf{x}}_{k+1} &= \hat{\mathbf{x}}_k - \eta_k \gamma_k \nabla f(\hat{\mathbf{x}}_k) + \gamma_k \hat{\mathbf{p}}_k,\\
\hat{\mathbf{p}}_{k+1} &= \hat{\mathbf{p}}_k - \eta_k  \nabla f(\hat{\mathbf{x}}_k),
\end{align}
where $\hat{\mathbf{x}} = (\hat{x}_1, \dots, \hat{x}_n)^T$ and $\hat{\mathbf{p}} = (\hat{p}_1, \dots, \hat{p}_n)^T$, and the subscript $k$ denotes the input quadrature at the $k$-th layer, while $k+1$ denotes the output of the $k$-th layer (and input to the next layer). This expression indicates that CV-QAOA can be interpreted as an iterative optimization procedure analogous to gradient-based methods with momentum. Consequently, increasing the circuit depth enhances the expressibility of the variational ansatz and can improve optimization performance in the absence of noise. In realistic noisy settings, however, deeper circuits do not necessarily yield better results. Thus, effective layer-dependent parameter schedules, such as decreasing $\gamma_k$ and increasing $\eta_k$ as the layer index $k$ increases, become important.

\subsection{Operations on the quad-rail lattice cluster state}
\label{sec:QRL}
The QRL cluster state is a two-dimensional entangled resource that enables MBQC in CV quantum systems~\cite{menicucci2011temporal}. In particular, by employing time-domain multiplexing, it is possible to efficiently generate large-scale cluster states with a large number of modes. The QRL cluster state generated in the time domain is composed of macronodes, each of which consists of four quantum modes referred to as micronodes. Each micronode at time step $k$ is entangled with micronodes at different time steps, $k-N$, $k-1$, $k+1$ and $k+N$. Here $N$ is determined by the size of QRL and corresponds to the maximum number of inputs. Quantum information processing is realized by sequentially performing measurements on the macronodes.

At time step $k$, two of the four micronodes in a macronode correspond to input states, while the remaining two micronodes are entangled with micronodes at other time steps ($k+1$ and $k+N$) via Einstein-Podolsky-Rosen (EPR) correlations. By performing collective measurements on the four micronodes at time step $k$ and applying feedforward operations to the entangled modes at time step $k+1$ and $k+N$ based on the measurement outcomes, quantum operations on the input states are implemented. As a result, the two input modes are processed to two output modes at subsequent macronodes, and quantum information propagates sequentially along the time direction. For further details of the operations on the QRL cluster state, see Refs.~\cite{alexander2016flexible,yokoyama2025full}.

In order to implement CV-QAOA, we employ the Gaussian operations that can be realized on macronodes. Those operations include the controlled-Z gate, $\hat{C}_Z(g) = e^{i g \hat{x}_1 \hat{x}_2}$, the X-invariant shear, $\hat{P}(\kappa) = e^{i \kappa \hat{x}^2}$, and the P-invariant shear $\hat{Q}(\mu) = e^{i \mu \hat{p}^2}$. It is noted that the single mode operations $\hat P$ and $\hat Q$ are a special case of two-input two-output macronode operations.

The Gaussian operations can be described in the Heisenberg picture as
\begin{align}
\hat{\bm{\xi}}_{\text{out}} = \bm{S} \hat{\bm{\xi}}_{\text{in}} + \hat{\bm{\Delta}},
\end{align}
where $\hat{\bm{\xi}}^T = (\hat{\mathbf{x}}^T, \hat{\mathbf{p}}^T)$, $\bm{S}$ is a symplectic matrix representing the Gaussian transformation, and $\hat{\bm{\Delta}}$ is a zero-mean noise operator arising from finite squeezing of the QRL cluster states, characterized by an uncorrelated covariance matrix $\mathrm{Cov}(\hat{\bm{\Delta}}) = e^{-2r_{\text{QRL}}} \bm{I}$. Furthermore, the $p$-displacement $\hat{D}(\nu) = e^{i \nu \hat{x}}$, which is necessary to implement CV-QAOA, is equivalently implemented by adding an offset through feedforward processing based on the measurement outcomes.

Input states to CV-QAOA are prepared through an initialization process. In this process, we measure one mode of the EPR pair that composes the QRL cluster state. By applying feedforward, we can prepare a squeezed state at the remaining mode, which serves as an input mode. This process, however, inevitably adds noise due to finite-resource squeezing.

\begin{figure*}[t!]\centering
\includegraphics[scale=1.0]{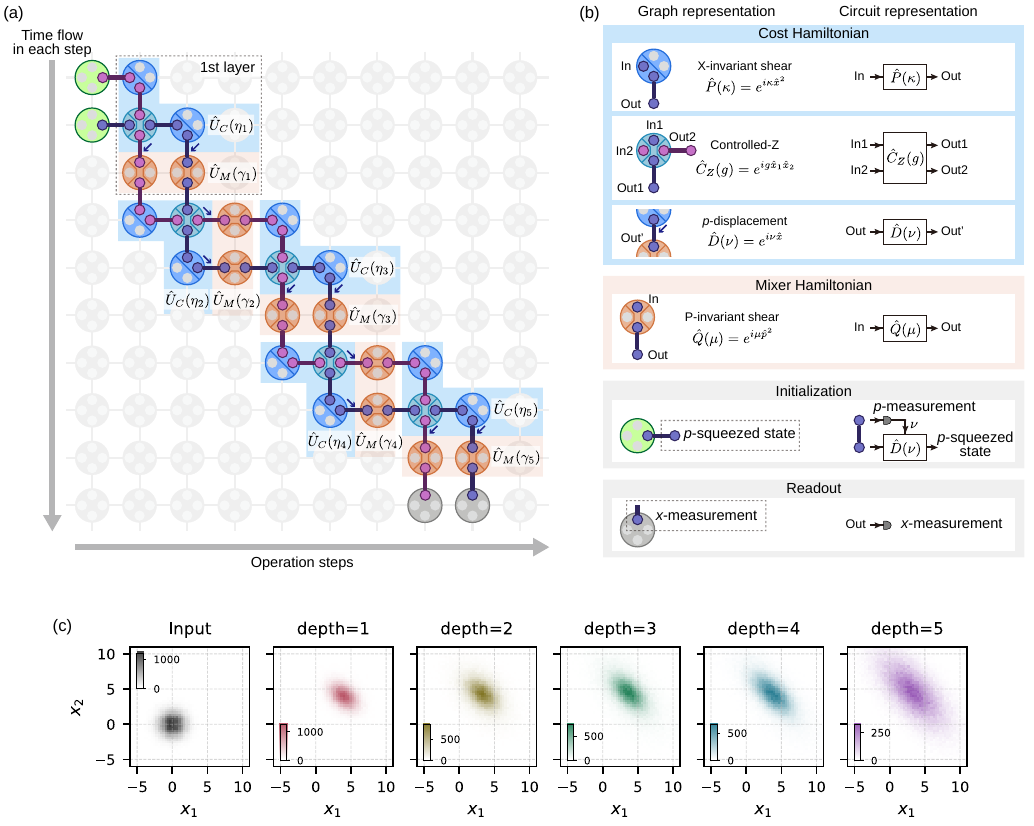}
\caption{
CV-QAOA implementation on the QRL cluster state.
(a) Example of the CV-QAOA ansatz mapped onto the QRL cluster state for a two-variable cost function with QAOA depth $d=5$.
(b) Basic measurement-based operations on a macronode, including Gaussian operations, state preparation, and measurement.
(c) Measured output quadrature distributions for the two-variable problem with optimized parameters. The histograms are constructed from $10^5$ measurement shots. Each circuit uses the optimized parameters corresponding to the best-performing points shown in \figref{4} (see main text for details).
}
\label{fig:2}
\end{figure*}

\section{Experimental realization of CV-QAOA}
\label{sec:setup}
\subsection{CV-QAOA ansatz on a QRL cluster state}
In this section, we describe how the CV-QAOA introduced in the previous section can be implemented on the QRL cluster state. We consider a quadratic cost function for continuous variables, given by
\begin{align}
    \label{eq:cost_function}
f(\mathbf{x}) = \frac{1}{2}\mathbf{x}^T \bm{A} \mathbf{x} + \mathbf{b}^T \mathbf{x} + c,
\end{align}
where the constant term $c$ does not affect the unitary implementation and is therefore omitted in the following discussion, although it is retained when evaluating the cost function.

In general, an arbitrary Gaussian transformation can be implemented on the QRL using a Bloch–Messiah decomposition, which consists of beam splitter networks and squeezing operations~\cite{yoshikawa2025configuration}. However, such implementations typically require a large number of macronodes, leading to unavoidable noise accumulation. In this work, we exploit the fact that the cost function depends only on position quadratures, allowing us to decompose the cost Hamiltonian into single-macronode Gaussian operations, directly map them on the QRL, and thereby achieve a more efficient implementation~\cite{yoshikawa2025configuration}.

The unitary corresponding to the cost Hamiltonian $\hat H_C=f(\hat{\textbf{x}})$ can be expressed as
\begin{align}
    \notag
\hat U_C(\eta)
&=
\left(\prod_i \hat P^{(i)}\!\left(-\tfrac{1}{2}\eta A_{i,i}\right)\right)
\left(\prod_{i<j} \hat C_Z^{(i,j)}(-\eta A_{i,j})\right)\\
&\quad \times
\left(\prod_i \hat D^{(i)}(-\eta b_i)\right),
\label{eq:cost_unitary}
\end{align}
where the superscripts indicate the modes on which the operators act. Here, the cost Hamiltonian is decomposed using the X-invariant shear $\hat P$ for the $\hat x_i^2$ terms, the controlled-Z $\hat C_Z$ for the $\hat x_i \hat x_j$ terms, and $p$-displacement $\hat D$ for the $\hat x_i$ terms. Each of these operations can be implemented using a single macronode. Since all of these operators are functions of $\hat{\textbf{x}}$, they commute with one another, enabling an efficient implementation on the QRL cluster state.

In addition, the mixer Hamiltonian is applied as P-invariant shear operations on each mode, given by
\begin{align}
\hat{U}_M(\gamma) = \prod_i \hat{Q}^{(i)}(-\tfrac{1}{2}\gamma).
\end{align}

\figureref{2}(a) shows an example of a two-variable problem with depth $d=5$, while the correspondence between the graph and circuit representations is summarized in \figref{2}(b). In this representation, each layer consists of a cost Hamiltonian block, in which two X-invariant shear macronodes and one controlled-Z macronode are arranged in a triangle and followed by the $p$-displacements, and a mixer Hamiltonian block containing two P-invariant shear macronodes. For each successive layer, the orientation of this block changes. The initial $p$-squeezed states are prepared at the upper-left corner via the initialization process. Then quantum information propagates along a zigzag path, proceeding toward the lower-right. Finally, the output quadratures are obtained at the measurement macronodes at the lower-right corner.

\figureref{2}(c) shows representative output quadrature distributions for a two-variable cost function implemented according to this mapping (see Sec.~\ref{sec:results} for the corresponding quantitative results). As the depth increases, the mean values tend to move toward the optimal solution $(3,5)$, while the distributions become broader because of accumulated noise. The figure is intended to provide an intuitive picture of how the output distribution changes with depth.

The implementation on the QRL cluster state can be systematically extended to problems with a larger number of variables. An $n$-variable cost function requires $n$ X-invariant shear operations and $n(n-1)/2$ controlled-Z operations from Eq.~\eqref{eq:cost_unitary}. These operations are efficiently arranged using $n(n+1)/2$ macronodes in an isosceles right-triangle pattern.  The X-invariant shear operation macronodes are placed along the hypotenuse of this triangle, while the remaining macronodes correspond to the controlled-Z operations. The $p$-displacements are then applied to the output of this pattern, followed by $n$ mixer P-invariant shear operation macronodes. In this way, problems of arbitrary size can be efficiently implemented in the QRL cluster state.

\subsection{Experimental setup and conditions}
In this work, we implement CV-QAOA using a cloud-based system capable of performing MBQC on the QRL cluster state, whose performance has been characterized in Ref.~\cite{yokoyama2025full}. In this system, a large-scale QRL cluster state is generated via time-domain multiplexing, enabling parallel processing of up to $N=101$ modes. The temporal duration of a single macronode is \SI{10}{\nano\second}, corresponding to a clock frequency of \SI{100}{\mega\hertz}. The experiment is controlled via an open-source Python SDK~\cite{mqc3SDK}, where jobs are constructed by embedding the desired operations (including their parameters) for each macronode. The measurement basis switching and data acquisition for each macronode are performed in real time by a field-programmable gate array (FPGA). By specifying the number of samples $n_{\mathrm{shots}}$ for each job and submitting it through Amazon Web Services (AWS), the corresponding measurement outcomes can be obtained.

A single measurement shot is set to be \SI{0.15}{\milli\second} long, including the inherent dead time due to system phase-locking. The total acquisition time is calculated as the single shot duration multiplied by the number of shots ($n_{\mathrm{shots}}$), plus the additional data transfer time from the FPGA to the server and communication latency in the cloud.

In the experiments reported in this work, the squeezing level of the QRL cluster state resource (corresponding to $e^{-2r_{\mathrm{QRL}}}$) is approximately \SI{-4.5}{\decibel}, while the squeezing and anti-squeezing levels of the input states  prepared by the initialization process are $\mathrm{Var}[\hat{p}_{\mathrm{in}}] \simeq \SI{-0.9}{\decibel}$ and $\mathrm{Var}[\hat{x}_{\mathrm{in}}] \simeq \SI{4.3}{\decibel}$, respectively.

\subsection{Classical optimization procedure}
The variational parameters are optimized on a classical computer based on measurement outcomes obtained from the quantum ansatz.
Specifically, samples $\hat{\mathbf{x}}$ are acquired via a cloud-based interface, and the expectation value of the cost function, $\langle f(\hat{\mathbf{x}}) \rangle$, is evaluated on a local computer.
The variational parameters are then iteratively updated based on this evaluation.

The estimation of the cost function is subject to statistical noise due to finite sampling, and achieving high accuracy and precision require a large number of measurements, resulting in a high evaluation cost.
Moreover, under such conditions, gradient estimation becomes unreliable.
For this reason, we employ a gradient-free optimization method, namely Bayesian optimization (BO), which is well suited for noisy black-box functions and enables efficient exploration with a limited number of evaluations~\cite{shahriari2015taking}. This choice is also consistent with previous CV photonic variational experiments~\cite{enomoto2023continuous,nielsen2025variational}. The optimization is implemented using the Python library Optuna~\cite{akiba2019optuna}, where the search is guided by a Gaussian process model.

The objective function is defined as the logarithm of the expectation value of the cost function, $\log_{10}\langle f(\hat{\mathbf{x}}) \rangle$.
In addition, the variational parameters $\{\eta_k, \gamma_k\}$ are optimized in the logarithmic domain.
This choice mitigates scale dependence and improves numerical stability when the relevant quantities span several orders of magnitude.
The parameters are restricted to the range $\eta_k, \gamma_k \in [10^{-2}, 10^{2}]$, taking into account experimentally accessible control precision.

For circuit depths greater than one, we impose monotonic constraints on the parameters, $\eta_k \le \eta_{k+1}$ and $\gamma_k \ge \gamma_{k+1}$.
This heuristic choice is motivated by empirical observations in qubit-based QAOA,  where optimized parameters often exhibit regular layer-dependent structures~\cite{akshay2021parameter, zhou2020quantum, he2023alignment}.
Such constraints reduce the effective search space and help stabilize the optimization.

The optimization is initialized from randomly sampled parameter sets and subsequently updated according to the BO procedure.
Since the optimization process is inherently stochastic due to both random initialization and measurement noise, multiple independent runs are performed under identical conditions, and statistical properties are evaluated over these repetitions.
Further details of the BO settings, as well as the specific numbers of trials used in each case, are provided in the Appendix~\ref{sec:BO}.

\section{Experimental results}
\label{sec:results}
\subsection{Two-variable demonstration}

\begin{figure*}[t!]\centering
\includegraphics[scale=0.9]{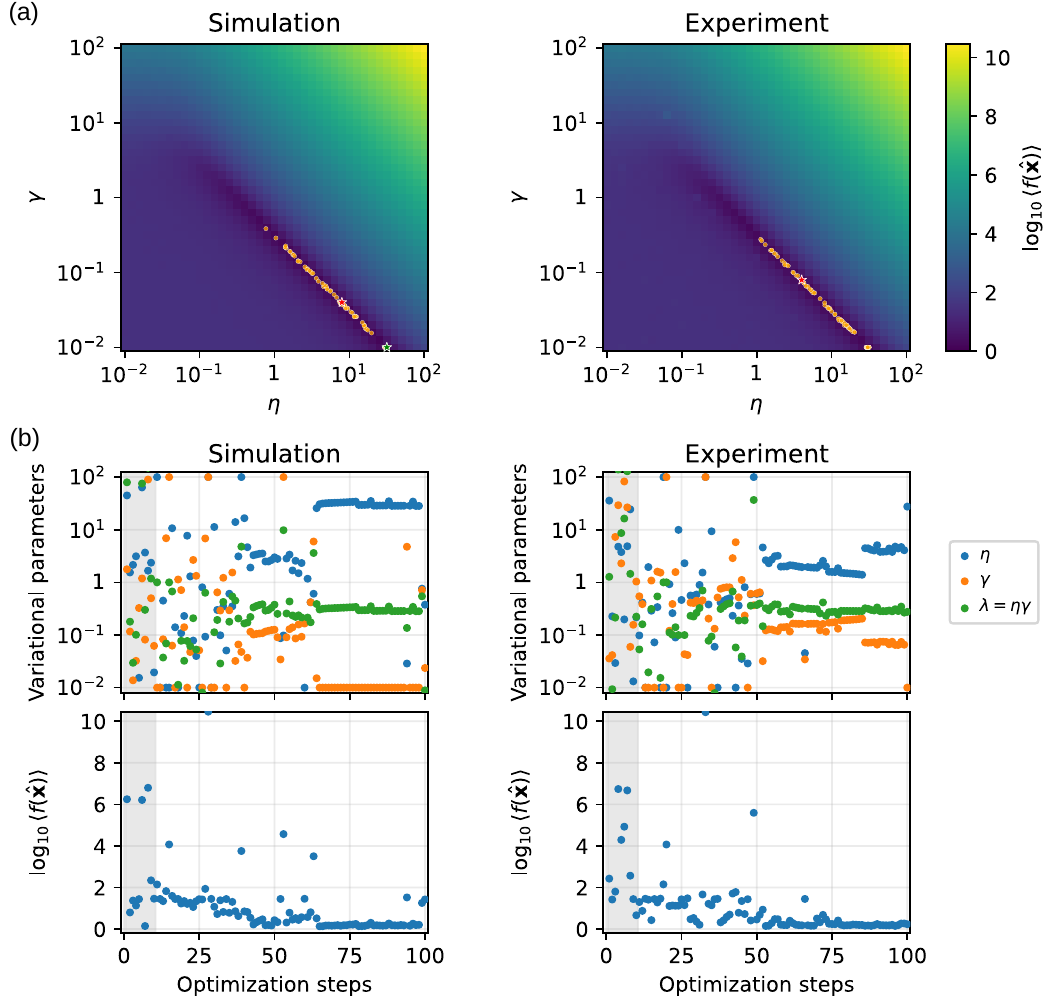}
\caption{
Cost function landscape and BO process for the two-variable problem at depth $d=1$.
(a) Experimental and simulated cost function landscape evaluated on a $41 \times 41$ grid over the variational parameters $(\eta,\gamma)$, where each point corresponds to an expectation value estimated from 1,000 measurement shots.
The red star indicates the optimal parameter obtained by grid search, while the green star in the simulation corresponds to the optimum in the absence of sampling noise.
The orange circles show the optimal parameters obtained from 100 independent runs of BO.
(b) Representative BO trajectory over 100 classical optimization steps.
The grey region indicates the initial random sampling stage with $n_{\mathrm{init}}=10$.
}
\label{fig:3}
\end{figure*}

We first demonstrate the operation of CV-QAOA using the following two-variable quadratic function:
$f(\mathbf{x}) = 1.2x_1^2 + x_2^2 + 1.4x_1x_2 - 14.2x_1 - 14.2x_2 + 56.8$. This function attains its minimum value of 0 at $(x_1, x_2) = (3,5)$.
\figureref{3} summarizes the experimental and simulated optimization behavior for this two-variable problem at depth $d=1$.
\figureref{3}(a) shows the cost function landscape evaluated over a grid of variational parameters $(\eta,\gamma)$.
In addition to the grid search results, the figure also displays the optimal parameters obtained from 100 independent runs of BO.
\figureref{3}(b) shows a representative BO trajectory over 100 classical optimization steps.
The first $n_{\rm init}=10$ steps correspond to random sampling, after which the BO procedure updates the parameters by balancing exploration of the parameter space and exploitation of regions expected to yield lower cost values.
To be consistent with the experimental conditions, the simulation parameters are set to a resource squeezing level of \SI{-4.5}{\decibel} and input squeezing and anti-squeezing levels of \SI{-0.9}{\decibel} and \SI{4.3}{\decibel}, respectively.

The resulting landscape exhibits a valley structure along the direction where $\eta\gamma$ is constant, indicating the presence of a flat direction with a small gradient near the minimum. Consistent with this structure, the optimal parameters obtained via BO do not converge to a single point but are broadly distributed along the valley floor, while maintaining an approximately constant product $\eta\gamma$.
This behavior is consistent with the theoretically predicted dependence on the product $\eta\gamma$. Furthermore, the analytically obtained optimal solution (green star in simulation) corresponds to the limit $\gamma \to 0$, in agreement with the analysis presented in the Appendix~\ref{sec:noise}.

On the other hand, both the BO and grid search results exhibit noticeable fluctuations. These variations arise from statistical noise due to the finite number of measurement shots, indicating that accurate estimation of the expectation value is required in flat regions of the landscape. Nevertheless, the experimental results are in good overall agreement with the simulations, confirming that the proposed implementation operates as expected.

\begin{figure}[t!]\centering
\includegraphics[width=\columnwidth]{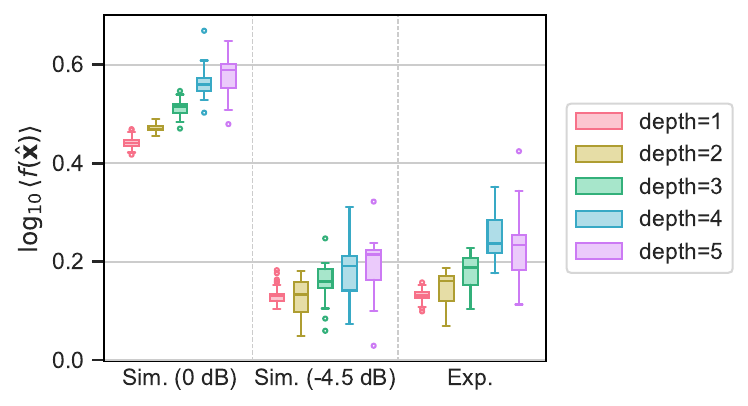}
\caption{
Optimization results obtained by BO as a function of depth.
Results are shown for both the classical benchmark (\SI{0}{\decibel} resource squeezing) and the quantum case with resource squeezing of \SI{-4.5}{\decibel}.
Each box plot represents the distribution of optimized cost values over multiple independent optimization runs (100 for depth $d=1$, 20 for depths $d>1$).
}
\label{fig:4}
\end{figure}

\figureref{4} shows the optimization results as a function of circuit depth from $d=1$ to $5$.
For comparison, results obtained without resource squeezing (\SI{0}{\decibel}) are included as a classical benchmark.
In this case, increasing the depth does not improve the cost function value, but instead leads to a monotonic degradation.
In contrast, when quantum resources are employed, a tendency toward performance improvement with increasing depth is observed, although the effect remains limited.
This behavior can be understood from the output distributions shown in \figref{2}(c), where the optimized parameters shift the mean toward the optimum, while the distribution becomes broader at larger depths due to accumulated noise.
At the same time, the spread of the results increases with depth, indicating a reduction in optimization stability.
This behavior can be attributed to both the difficulty of BO and the accumulation of noise in deeper circuits.
Nevertheless, the experimental results remain in good overall agreement with the simulations, confirming the validity of the proposed implementation.

\subsection{Scaling to higher dimensions}
\begin{figure*}[b!]\centering
\includegraphics[scale=1.0]{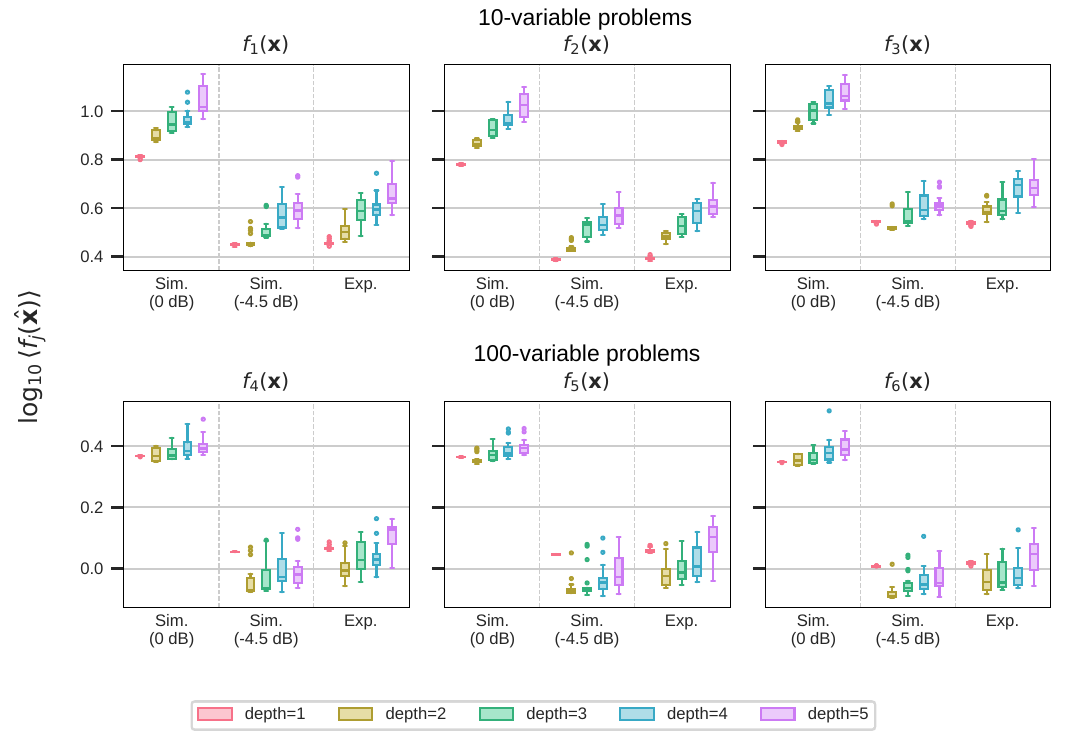}
\caption{
Optimization results for randomly generated quadratic problems with 10 and 100 variables.
The functions $f_1(\mathbf{x})$--$f_3(\mathbf{x})$ correspond to 10-variable problems, while $f_4(\mathbf{x})$--$f_6(\mathbf{x})$ correspond to 100-variable problems.
For the 100-variable case, the cost matrix $\bm{A}$ is diagonal.
Each box plot represents the distribution of optimized cost values obtained from multiple independent BO runs (20 attempts for all depths).
}
\label{fig:5}
\end{figure*}

Since our platform can handle a large number of modes, we investigate the scalability of CV-QAOA to higher-dimensional optimization problems.
\figureref{5} shows the results for randomly generated problems with 10 and 100 variables.
All problem instances are generated and normalized as described in the Appendices~\ref{sec:random_cost} and~\ref{sec:normalization}.
For the 100-variable case, the cost matrix $\bm{A}$ is taken to be diagonal so that each mode evolves independently, and the CV-QAOA implementation effectively reduces to a set of parallel single-variable problems.
The corresponding QRL graph representations for the 10- and 100-variable implementations are shown in Appendix~\ref{sec:qrl_graphs}.

For the 10-variable problems, we observe behavior similar to the two-variable case, where performance improvements with increasing depth are limited and tend to saturate around depth $d=2$.
In contrast, for the 100-variable problems, the improvement from depth $d=1$ to $2$ is more pronounced, although further increases in depth yield only limited gains.

\subsection{Intrinsic performance under idealized optimization}
The results presented so far are based on BO and therefore include the effects of optimization performance and statistical noise. As such, they do not directly reflect the intrinsic performance of the CV-QAOA itself. To isolate the algorithmic capability, we instead evaluate the performance under idealized conditions, i.e., using gradient-based optimization with an infinite number of shots.

\begin{figure*}[t!]\centering
\includegraphics[scale=1.0]{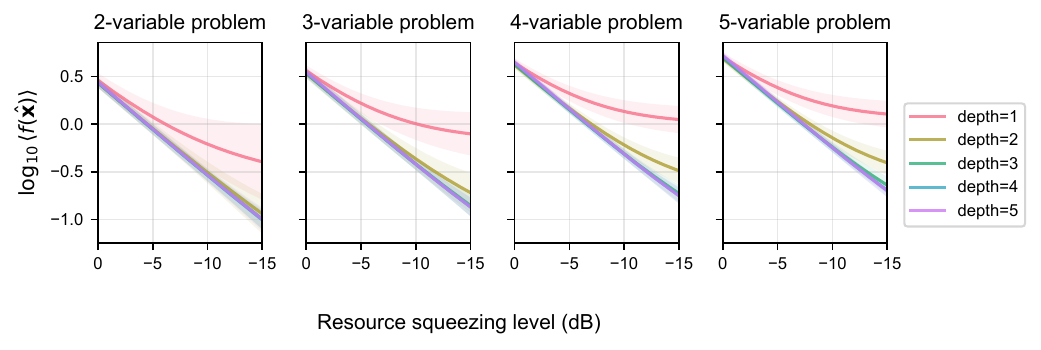}
\caption{
Performance of CV-QAOA evaluated under the idealized conditions using gradient-based optimization with an infinite number of shots.
The results show the mean and standard deviation of the optimized cost values over multiple randomly generated problem instances for 2- to 5-variable problems.
Sampling noise is removed by assuming an infinite number of measurement shots ($n_{\mathrm{shots}}=\infty$).
The initial state is taken to have $(-10,\,10)$ dB squeezing and anti-squeezing.
}
\label{fig:6}
\end{figure*}

\figureref{6} shows the optimized cost values of CV-QAOA under the idealized conditions, evaluated across multiple randomly generated problem instances. For each instance, the variational parameters are optimized until convergence.

From these results, we observe that the performance of CV-QAOA improves with increasing circuit depth. At the same time, the required depth to achieve comparable performance increases with the problem dimension. In addition, larger resource squeezing leads to better performance, reflecting the reduction of noise and improved convergence toward the optimal state.

In contrast to the BO-based results, these findings indicate that CV-QAOA is, in principle, capable of achieving improved performance with increasing depth. The limitations observed in the experimental results therefore primarily originate from the difficulty of classical optimization and the presence of noise, rather than from a fundamental limitation of the algorithm itself.

\section{Conclusion}
\label{sec:conclusion}

In this work, we implemented the CV-QAOA ansatz on a QRL cluster state and experimentally investigated its performance on quadratic optimization problems using BO. We also presented a systematic mapping of arbitrary quadratic cost functions onto the QRL architecture. This mapping is, in principle, scalable to large numbers of modes, and we examined its experimental behavior for multi-variable problems at multiple QAOA depths.

Experimentally, we found that increasing the depth from 1 to 2 leads to a modest improvement in performance, whereas further increases yield only limited gains. For the 2- and 10-variable problems studied here, the benefit of deeper circuits was limited under realistic experimental conditions. This behavior can be attributed to two main factors: the accumulation of noise with increasing circuit size, and the growth of statistical uncertainty under finite-shot measurements, which makes the classical optimization more difficult.

In contrast, numerical analysis under idealized conditions, i.e., assuming an infinite number of measurement shots and gradient-based optimization, shows that the performance of CV-QAOA can improve with increasing depth. This comparison indicates that the experimentally observed limitations arise primarily from noise and classical optimization difficulty, rather than from a fundamental limitation of the CV-QAOA ansatz itself. In particular, for the class of problems considered here, the cost landscape exhibits a flat structure near the optimum, which is unfavorable for black-box optimization methods such as BO.

Several directions remain for future work. On the algorithmic side, it will be important to develop optimization strategies better suited to such flat landscapes, including parameter-transfer and initialization methods~\cite{zhou2020quantum,sack2021quantum}, layerwise training strategies~\cite{campos2021training}, and gradient-evaluation methods adapted to CV systems~\cite{schuld2019evaluating,nielsen2025variational}. On the implementation side, extending the framework beyond quadratic cost functions is an important direction. Non-Gaussian operation provides a route to higher-order polynomial cost terms~\cite{lloyd1999quantum,gottesman2001encoding}. In the QRL architecture, the cubic phase operation associated with Hamiltonians proportional to $\hat{x}_i^3$ can be incorporated using a non-Gaussian ancillary state and nonlinear feedforward~\cite{yokoyama2025full,sakaguchi2023nonlinear}. Explicit and scalable decompositions for general multimode higher-order cost Hamiltonians, as well as more efficient mappings with reduced noise accumulation, remain to be developed. Although quantum advantage is not expected in the present Gaussian and quadratic setting, this work nevertheless provides one of the first experimental demonstrations of CV-QAOA on a programmable CV quantum platform and offers a concrete foundation for future studies of variational quantum algorithms in continuous-variable systems.

\section*{ACKNOWLEDGMENTS}
This work was supported by the Japan Science and Technology (JST) Agency (Moonshot R \& D) Grant No. JPMJMS2064 and JPMJMS256I, the UTokyo Foundation, and donations from Nichia Corporation of Japan.
H.N. acknowledges financial support from The Forefront Physics and Mathematics Program to Drive Transformation (FoPM). H.N. and W.A. acknowledge funding from the Japan Society for the Promotion of Science KAKENHI (No. 23K13040, 24KJ0745).

\section*{DATA AVAILABILITY}
The experimental data in this article are available upon reasonable request from the authors.

\appendix
\section{Generation of random problem instances}
\label{sec:random_cost}

In this section, we describe how random quadratic optimization problems are prepared.
We generate cost functions of the form
$f(\mathbf{x}) = \frac{1}{2}\mathbf{x}^T \bm{A} \mathbf{x} + \mathbf{b}^T \mathbf{x}$,
where $\bm{A}$ is a symmetric positive semidefinite matrix.

For each problem instance, we first generate a random matrix $\bm{M} \in \mathbb{R}^{n \times n}$
whose entries are independently sampled from a uniform distribution
$M_{ij} \sim \mathcal{U}[-\sqrt{3/n},\, \sqrt{3/n}]$,
and construct $\bm{A} = 2\bm{M}\bm{M}^T$, which ensures that $\bm{A}$ is symmetric and positive semidefinite.
For the optimization problems considered here, we restrict to the nonsingular case.

For high-dimensional cases ($n=100$), we instead use a diagonal matrix
$\bm{A} = \mathrm{diag}(a_1,\dots,a_n)$ with $a_i \sim \mathcal{U}[1,5]$
to simplify the implementation of the ansatz, effectively corresponding to a set of parallel single-variable problems.

We then sample a random vector $\mathbf{z} \sim \mathcal{U}([-3,3]^n)$ and define $\mathbf{b} = -\bm{A}\mathbf{z}$,
so that $\mathbf{z}$ corresponds to the global minimizer $\mathbf{x}_{\min}$ when $\bm{A}$ is nonsingular.
For \figref{5}, we generate three random instances for each dimension ($n=10$ and $n=100$),
where dense matrices are used for $n=10$ and diagonal matrices for $n=100$.
For \figref{6}, we generate $100$ random instances for each dimension $n=2,3,4,5$ using the same procedure.

\section{Normalized cost function}
\label{sec:normalization}

To remove scale dependence across problem instances and enable comparison between cost functions with different dimensions and coefficients, we introduce the normalized cost function.
For the quadratic cost function $f(\mathbf{x}) = \frac{1}{2}\mathbf{x}^T \bm{A} \mathbf{x} + \mathbf{b}^T \mathbf{x}$, we define the normalized cost function as
\begin{align}
\tilde f(\mathbf{x}) = \frac{f(\mathbf{x}) - f(\mathbf{x}_{\min})}{\mathrm{Tr}[\bm{A}/2]}.
\end{align}
Here, $\mathbf{x}_{\min}$ denotes the minimizer of the cost function.
By construction, $\tilde f(\mathbf{x}_{\min}) = 0$, and the resulting cost values are rescaled to suppress dependence on both the overall scale of $\bm{A}$ and the problem dimension.
The normalization factor $\mathrm{Tr}[\bm{A}/2]$ reflects the effective scale of the quadratic term.

This normalization can be interpreted as scaling the cost function by the typical magnitude of fluctuations in the measurement outcomes, which have a finite variance.
However, as the problem dimension $n$ increases, the overall scale of the coefficients in the normalized cost function becomes excessively small.
To compensate for this effect in the implementation of the cost Hamiltonian within the ansatz, we further introduce a rescaled cost function, $n \tilde f(\mathbf{x})$, so that the cost values remain at a comparable scale across different dimensions.

\section{Classical optimization details}
\label{sec:BO}
The BO is implemented using the Optuna library~\cite{akiba2019optuna} with a Gaussian process-based sampler.
The search is guided by an acquisition function based on expected improvement.
The Gaussian process model employs a Matérn kernel ($\nu = 2.5$), following the default implementation of the Optuna Gaussian process sampler.
For depths greater than one, monotonic constraints on the parameters are incorporated into the sampler, enabling constrained optimization.

The optimization is characterized by three parameters: the number of initial random samples $n_{\mathrm{init}}$, the total number of trials $n_{\mathrm{trial}}$, and the number of independent optimization runs $n_{\mathrm{attempt}}$.
The initial samples are generated randomly, and subsequent trials are determined sequentially according to the acquisition function.
Each evaluation of the cost function is based on $n_{\mathrm{shots}} = 1,000$ measurement samples.

The specific settings used in each figure are as follows.
For \figref{3}, we use $n_{\mathrm{init}}=10$, $n_{\mathrm{trial}}=100$, and $n_{\mathrm{attempt}}=100$.
For \figref{4}, we use $n_{\mathrm{init}}=10$, $n_{\mathrm{trial}}=100$, and $n_{\mathrm{attempt}}=100$ for depth $d=1$, and $n_{\mathrm{init}}=30$, $n_{\mathrm{trial}}=200$, and $n_{\mathrm{attempt}}=20$ for depths $d>1$.
For \figref{5}, we use $n_{\mathrm{init}}=10$, $n_{\mathrm{trial}}=100$, and $n_{\mathrm{attempt}}=20$ for depth $d=1$, and $n_{\mathrm{init}}=30$, $n_{\mathrm{trial}}=200$, and $n_{\mathrm{attempt}}=20$ for depths $d>1$.

\begin{figure*}[t !]\centering
\includegraphics[width=\textwidth]{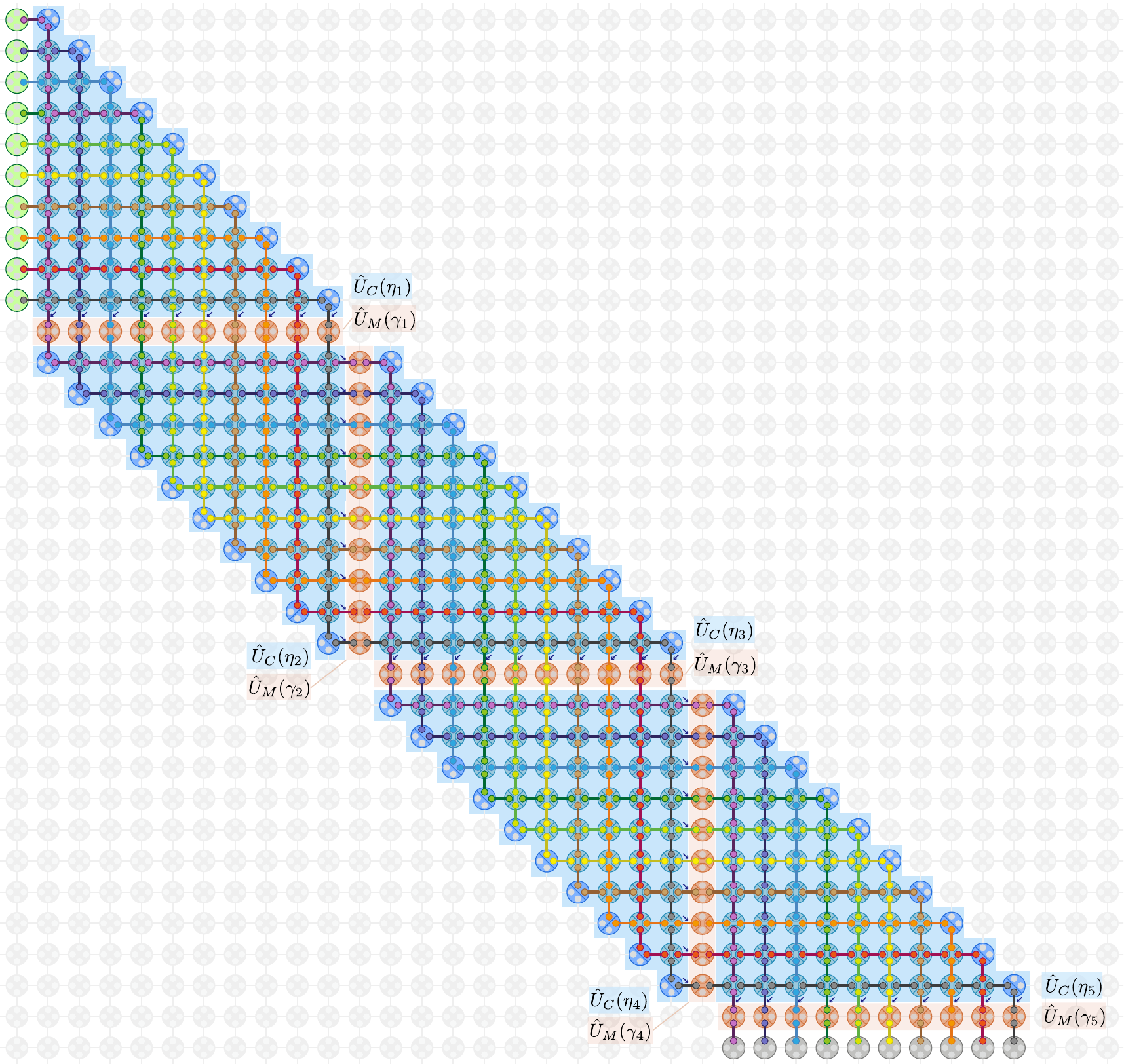}
\caption{
QRL graph representation of the experimentally implemented CV-QAOA ansatz for 10-variable quadratic problems at depth $d=5$.
}
\label{fig:10_variables}
\end{figure*}

\section{Noise analysis of QAOA on the QRL cluster state}
\label{sec:noise}
In this section, we explicitly describe the input--output relations of the CV-QAOA ansatz at each layer for a quadratic cost function with $n$ variables.
We consider the implementation described in the main text, where $n$ input modes enter the $k$-th layer and $n$ output modes are obtained after the mixer operation, and express the noise at the $(k+1)$-th layer in terms of the variables and noise contributions at the $k$-th layer.
To this end, we label the macronodes by a pair of indices $(i,j)$, where $i$ and $j$ correspond to the vertical and horizontal directions, respectively.
Without loss of generality, the input macronodes are labeled as $(i,1)$ for $i=1,\dots,n$, and the subsequent cost and mixer Hamiltonian macronodes are labeled accordingly.
We denote the noise operators affecting the quadratures of mode $l$ at macronode $(i,j)$ by $\hat{\delta}^{(i,j)}_{x,l}$ and $\hat{\delta}^{(i,j)}_{p,l}$.

The input--output relations at each macronode for the cost Hamiltonian are given as follows:
\begin{itemize}

\item \textbf{Macronode $(i,i)$ ($i=1,\dots,n$): X-invariant shear}
\begin{align}
    \hat x_i &\to \hat x_i + \hat{\delta}^{(i,i)}_{x,i}, \\
    \hat p_i &\to \hat p_i - \eta_k A_{i,i} \hat x_i + \hat{\delta}^{(i,i)}_{p,i}.
\end{align}

\item \textbf{Macronode $(i,j)$ ($1\le j<i\le n$): Controlled-Z}
\begin{align}
    \hat x_i &\to \hat x_i + \hat{\delta}^{(i,j)}_{x,i}, \\
    \hat p_i &\to \hat p_i - \eta_k A_{i,j} \hat x_j + \hat{\delta}^{(i,j)}_{p,i}, \\
    \hat x_j &\to \hat x_j + \hat{\delta}^{(i,j)}_{x,j}, \\
    \hat p_j &\to \hat p_j - \eta_k A_{i,j} \hat x_i + \hat{\delta}^{(i,j)}_{p,j}.
\end{align}

\item \textbf{Macronode $(n+1,j)$ ($j=1,\dots,n$): $p$-displacement and P-invariant shear}
\begin{align}
    \hat x_j &\to \hat x_j + \gamma_k (\hat p_j - \eta_k b_j) + \hat{\delta}^{(n+1,j)}_{x,j}, \\
    \hat p_j &\to \hat p_j - \eta_k b_j + \hat{\delta}^{(n+1,j)}_{p,j}.
\end{align}

\end{itemize}

From these relations, the input--output transformation for the cost Hamiltonian operations at the $k$-th layer can be written as
\begin{align}
    \hat{\mathbf{x}}' &= \hat{\mathbf{x}}_k + \hat{\bm{\Delta}}_{x,C}, \\
    \hat{\mathbf{p}}' &= \hat{\mathbf{p}}_k - \eta_k (\bm{A}\hat{\mathbf{x}}_k + \mathbf{b})
    + \hat{\bm{\Delta}}_{p,C} - \eta_k \hat{\bm{\Delta}}_{q,C},
\end{align}
where $\hat{\bm{\Delta}}_{x,C}$ and $\hat{\bm{\Delta}}_{p,C}$ denote accumulated noise in the quadratures, and $\hat{\bm{\Delta}}_{q,C}$ represents additional noise induced via the controlled-Z and shear operations.
These noise contributions can be written explicitly as
\begin{align}
    (\hat{\bm{\Delta}}_{x,C})_l
    &= \sum_{j=1}^{l-1} \hat{\delta}^{(l,j)}_{x,l}
     + \sum_{i=l}^{n} \hat{\delta}^{(i,l)}_{x,l}, \\
    (\hat{\bm{\Delta}}_{p,C})_l
    &= \sum_{j=1}^{l-1} \hat{\delta}^{(l,j)}_{p,l}
     + \sum_{i=l}^{n} \hat{\delta}^{(i,l)}_{p,l}, \\
    (\hat{\bm{\Delta}}_{q,C})_l
    &= \sum_{j=1}^{l-1} A_{l,j}
        \left( \sum_{m=1}^{j-1} \hat{\delta}^{(j,m)}_{x,j}
        + \sum_{i=j}^{l-1} \hat{\delta}^{(i,j)}_{x,j} \right) \notag \\
    &\quad + \sum_{i=l}^{n} A_{i,l} \sum_{j=1}^{l-1} \hat{\delta}^{(i,j)}_{x,i}.
\end{align}

Including the mixer Hamiltonian, the full input--output relation becomes
\begin{align}
\hat{\mathbf{x}}_{k+1}
& = \hat{\mathbf{x}}' +\gamma_k \hat{\mathbf{p}}' + \hat{\bm{\Delta}}_{x,M} \\
&= (\bm{I} - \lambda_k \bm{A}) \hat{\mathbf{x}}_k
   - \lambda_k \mathbf{b}
   + \gamma_k \hat{\mathbf{p}}_k \notag \\
&\quad + \gamma_k \hat{\bm{\Delta}}_{p,C}
   - \lambda_k \hat{\bm{\Delta}}_{q,C}
   + \hat{\bm{\Delta}}_{x,C}
   + \hat{\bm{\Delta}}_{x,M}, \\
\hat{\mathbf{p}}_{k+1}
&= \hat{\mathbf{p}}' + \hat{\bm{\Delta}}_{p,M}\\
&= \hat{\mathbf{p}}_k
   - \frac{\lambda_k}{\gamma_k}(\bm{A}\hat{\mathbf{x}}_k + \mathbf{b}) \notag \\
&\quad - \frac{\lambda_k}{\gamma_k} \hat{\bm{\Delta}}_{q,C}
   + \hat{\bm{\Delta}}_{p,C}
   + \hat{\bm{\Delta}}_{p,M},
\end{align}
where $\lambda_k = \eta_k \gamma_k$, and
\begin{align}
    (\hat{\bm{\Delta}}_{x,M})_l &= \hat{\delta}^{(n+1,l)}_{x,l}, \\
    (\hat{\bm{\Delta}}_{p,M})_l &= \hat{\delta}^{(n+1,l)}_{p,l}.
\end{align}

The expectation value of the cost function [Eq.~\eqref{eq:cost_function}] can be evaluated from the mean vector $\textbf{m}$ and covariance matrix $\bm{V}$ of the output quadratures $\hat{\mathbf{x}}_{\mathrm{out}}$ as
\begin{align}
\langle f(\hat{\mathbf{x}}_{\mathrm{out}}) \rangle
= \frac{1}{2}\mathrm{Tr}(\bm{A}\bm{V}) + \frac{1}{2}\textbf{m}^T \bm{A} \textbf{m} + \mathbf{b}^T \textbf{m} + c.
\end{align}

For depth $d=1$, the mean vector is given by
\begin{align}
\textbf{m} = -\lambda \mathbf{b},
\end{align}
where $\lambda = \eta\gamma$. We note that the variational parameters can be reparameterized in terms of $(\lambda, \gamma)$, and the mean therefore depends only on the product $\lambda$.

The dependence of the cost function on $\gamma$ arises solely through the covariance matrix, and differentiating with respect to $\gamma$ yields
\begin{align}
\frac{\partial \langle f(\hat{\mathbf{x}}_{\mathrm{out}})\rangle}{\partial \gamma}
&=
\frac{1}{2}\mathrm{Tr}\!\left(
\bm{A}\frac{\partial \bm{V}}{\partial \gamma}
\right) \\
&=
\gamma\,\mathrm{Tr}\!\left[
\bm{A}\left(
\mathrm{Cov}[\hat{\mathbf{p}}_{\mathrm{in}}]
+\mathrm{Cov}[\hat{\bm{\Delta}}_{p,C}]
\right)
\right].
\end{align}
Since the covariance matrices are positive semidefinite, this derivative is nonnegative for $\gamma>0$, and the optimal value is therefore attained at $\gamma\to 0$.

This implies that the optimization favors the limit $\gamma \to 0$, while the parameter that determines the mean displacement remains $\lambda = \eta\gamma$. Consequently, for a finite optimal value of $\lambda$, the corresponding optimal value of $\eta$ diverges. Therefore, at depth $d=1$, the variational parameters $\eta$ and $\gamma$ are not independent, and the optimization effectively reduces to a one-dimensional problem in terms of $\lambda$, with $\gamma$ constrained to the boundary $\gamma \to 0$.

In practice, the accessible ranges of $\eta$ and $\gamma$ are bounded in the experimental implementation, and the optimization is therefore constrained by these limits rather than reaching the asymptotic regime $\gamma \to 0$ and $\eta \to \infty$.

\section{Graph representations for 10- and 100-variable problems}
\label{sec:qrl_graphs}
\begin{figure}[htbp!]\centering
\includegraphics[height=0.95\textheight,keepaspectratio]{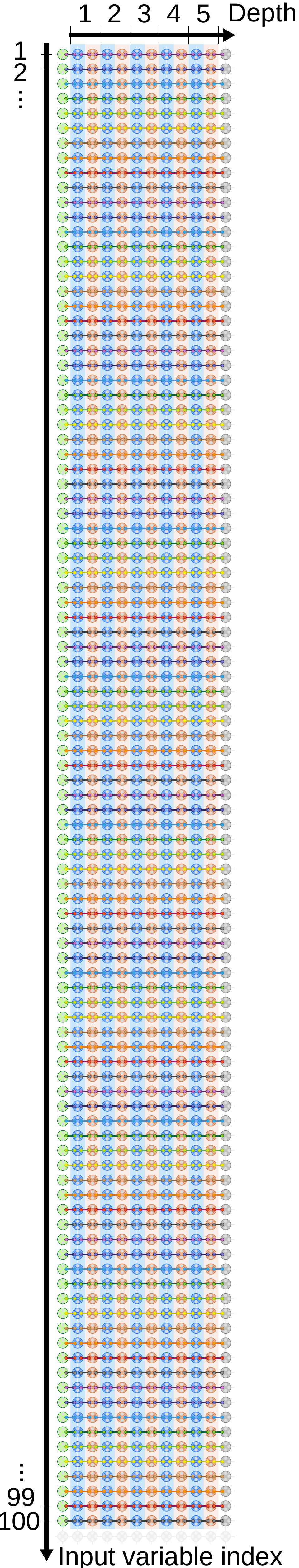}
\caption{
QRL graph representation of the experimentally implemented CV-QAOA ansatz for 100-variable quadratic problems at depth $d=5$.
}
\label{fig:100_variables}
\end{figure}

In this section, we show the graph representations of the experimentally implemented CV QAOA ansatz for 10- and 100-variable problems. For the 10-variable problem shown in \figref{10_variables}, we implement the cost Hamiltonian using a triangle-shaped set of macronodes, followed by a mixer Hamiltonian block composed of P-invariant shear macronodes. This sequence is repeated five times, corresponding to depth $d=5$. For the 100-variable problem in \figref{100_variables}, since we assume a diagonal quadratic function, the cost Hamiltonian is implemented by a single row of X-invariant shear macronodes followed by a row of mixer Hamiltonian macronodes (P-invariant shear). This sequence is repeated up to five times, corresponding to depth $d=5$.

\clearpage

%%%%%%%%%%%%%%%%%%%%%%%%%%%%%%%%%%%%%%%%%%%%%%%%
\bibliography{QAOA_Bib}

\end{document}